\documentclass{aa}
\usepackage{graphicx}
\usepackage{txfonts}
\usepackage{xspace}
\usepackage{comment}
\usepackage{amsmath}
\usepackage{gensymb}
\usepackage[colorlinks, citecolor={blue}]{hyperref}
\usepackage{multirow}
\usepackage{lscape}

\newcommand{\xmm}{XMM-\textit{Newton}\xspace}
\newcommand{\nustar}{\textit{NuSTAR}\xspace}

\newcommand{\mdot}{$\dot{M}$\xspace}
\newcommand{\mbh}{$M_\bullet$\xspace}
\newcommand{\msun}{$M_\odot$\xspace}
\newcommand{\fluxcgs}{erg cm$^{-2}$ s$^{-1}$\xspace}
\newcommand{\lumcgs}{erg s$^{-1}$\xspace}
\newcommand{\kerrbb}{\texttt{kerrbb}\xspace}

\begin{document} 

   \title{Toward measuring the spin of obscured supermassive black holes: a critical assessment with disk megamasers}

   \author{Alberto Masini
          \inst{1,2}
          \and
          Annalisa Celotti\inst{1,3,4,5}
          \and 
          Samuele Campitiello\inst{1}
          }

   \institute{SISSA - International School for Advanced Studies, Via Bonomea 265, 34151 Trieste, Italy \email{amasini@sissa.it} 
   \and 
   INAF - Osservatorio di Astrofisica e Scienza dello Spazio di Bologna, Via Gobetti 93/3, 40129 , Italy 
   \and
   INAF - Osservatorio Astronomico di Brera, Via E. Bianchi 46, I-23807, Merate, Italy
   \and
   INFN - National Institute for Nuclear Physics, Via Valerio 2, 34127 Trieste, Italy
   \and
   IFPU - Institute  for  Fundamental  Physics  of  the  Universe,  Via  Beirut  2, 34151 Trieste, Italy
   }
   \date{}
   
\abstract
   {Mass and spin are two fundamental properties of astrophysical black holes. While some established, indirect methods are adopted to measure both these properties of active galactic nuclei (AGN) when viewed relatively face-on, very few suggested methods exist to measure these properties when AGN are viewed highly inclined and potentially obscured by large amounts of gas.}
   {In this context, we explore the accuracy and performance of a recently proposed method to estimate the spin of AGN through fitting their accretion disk spectral energy distribution, when adapted for highly inclined and obscured systems, and in particular to a sample of six, local water megamasers. For these sources, both the accretion rate and inclination angle are known, allowing us to rely only on the AGN bolometric luminosity to infer their spin.}
   {Using the bolometric luminosity as a proxy for the accretion disk peak luminosity, we derive the expected bolometric luminosity as a function of spin. Then, we measure the bolometric luminosity of each source through X-ray spectroscopy, and compare it with the expected value to constrain the spin of the AGN.}
   {The quality of the constraints depend critically on the accuracy of the measured bolometric luminosity, which is difficult to estimate in heavily obscured systems. Three out of six sources do not show consistency between the expected and measured bolometric luminosities, while other three (four, when considering the [OIII] line as tracer of the bolometric luminosity) are formally consistent with high spin values.}
   {Our results suggest that this method, although promising (and possibly considered as a future calibrator for other methods) needs better observational data and further theoretical modeling to be successfully applied to obscured AGN and to infer robust results.}


   \titlerunning{SMBH spin in disk megamasers}
   \maketitle
   

\section{Introduction}

A few decades of research demonstrated that the mass \mbh of supermassive black holes (SMBHs) at the center of active galactic nuclei (AGN) is a crucial parameter in estimating the cosmic history of accretion \citep{soltan82, shankar09}, and that it correlates with the properties of their host galaxy bulges \citep{kormendyrichstone95,magorrian98,ferraresemerritt00,gebhardt00}. These correlations are considered the final product of the co-evolution of SMBHs and galaxies \citep{reinesvolonteri15}. 
The spin $a$, on the other hand, could also turn out to be of paramount importance to understand how SMBHs preferentially gain their mass through the cosmic history \citep{volonteri13,sesana14}. Indeed, if the growth of a SMBH happens primarily through sustained prolonged accretion of matter, the coherent angular momentum of the accretion disk would spin it up. Conversely, if the mass growth is dominated by a more chaotic and incoherent events, the final spin of the SMBH should be lower \citep{sesana14}. In addition, the rotational energy of a SMBH makes up a large reservoir that can be extracted \citep[for example through the so called BZ process,][]{blandfordznajek77} from its ergosphere, possibly powering the relativistic jets that are launched from the immediate vicinity of some AGN. 
Different observational methods are adopted to measure both the mass and spin of AGN when viewed relatively face-on. For instance, when the accretion disk and the broad line region are directly visible, the AGN mass and spin can be inferred through the reverberation mapping technique \citep[e.g.,][]{grier17} and the broadened shape of the iron $K\alpha$ line, respectively \citep[e.g.,][]{georgefabian91, reynolds19}. 
Recently, \citet{campitiello18} proposed another way to constrain the spin and mass of a relatively unobscured AGN through fitting its accretion disk emission in the optical-UV portion of the spectral energy distribution (SED), the big blue bump (BBB), with a relativistic accretion model \citep[\kerrbb,][]{li05}. In particular, the peak frequency and intensity of the BBB depend on a combination of four parameters: the inner accretion rate \mdot, BH mass \mbh, inclination angle of the system $\theta$, and BH spin $a$. However, a large fraction of AGN have their accretion disk and broad line region hidden by some amount of dusty obscuring gas, usually referred to as the torus \citep{merloni14, buchner15}. In these cases, these aforementioned methods cannot be applied, and new possibilities need to be searched and investigated. 

Disk megamasers, i.e. AGN with water maser emission at rest-frame 22 GHz tracing the Keplerian rotation of a sub-pc scale molecular disk orbiting the SMBH \citep[e.g.,][]{tarchi12}, are a subsample of local, heavily obscured AGN \citep{masini16} but for which exquisitely precise \mbh and inclination angles are known \citep[e.g.,][]{kuo11}. Therefore, they offer a unique possibility to apply the aforementioned SED fitting method, as two of the four unknown parameters are well determined. Indeed, disk megamasers opened a new and unique window on measuring BH masses for highly obscured and inclined systems \citep[and have been recently used to calibrate a general method for obscured AGN; see][]{gliozzi21}, although measuring or even constraining their spins has been so far prohibitive \citep[but see][for a recent, new suggestion to constrain their spin]{ginerloeb21}.

In this paper we adapt the SED fitting method of \citet{campitiello18} to explore its feasibility when applied to a few, selected megamasers for which an independent estimate of their inclination angle $\theta$, mass \mbh, accretion rate \mdot and of the bolometric luminosity $L_{\rm bol}$ exist or can be estimated from the literature. The paper is structured as follows: Section \ref{sec:methodology} presents the theoretical assumptions and lays the foundations of the method. Section \ref{sec:sample} presents the sample considered in this work. In Section \ref{sec:results} our results are presented, and commented in Section \ref{sec:discussion}. Finally, we draw our conclusions in Section \ref{sec:conclusions}. Appendix \ref{sec:appendix} presents the details about our own spectral analysis and measurement of the X-ray luminosities.
No cosmology is assumed since all distances are geometric, and uncertainties are quoted at $1\sigma$ confidence level, unless otherwise stated.

\begin{table*}
\caption{Sample considered in this work.}
\label{tab:info}
\centering                                    
\begin{tabular}{l c c c c c c c c c c}          
\hline\hline                  
Name & $D$ [Mpc] & $\theta$ [\degree] & Ref. & $\log$\mbh/$M_\odot$ & $R_{\rm in}$ [pc] & $R_{\rm out}$ [pc] & $R_{\rm infl}$ [pc] & $\log$\mdot/$M_\odot$ & $\log{L_{\rm bol}/\text{erg s$^{-1}$}}$ & $\log{\lambda_{\rm Edd}}$\\ 
(1) & (2) & (3) & (4) & (5) & (6) & (7) & (8) & (9) & (10) & (11) \\
\hline
    NGC 2960 & 81 & 89 & (a) & 7.09 & 0.13 & 0.31 & 1.92 & $-2.44^{+0.18}_{-0.28}$ & $43.11^{+0.24}_{-0.24}$ & $-2.1$ \\
    NGC 4258 & 7.6 & 72 & (b) & 7.60 & 0.11 & 0.30 & 12.98 & $-3.10^{+0.14}_{-0.20}$ & $41.78^{+0.07}_{-0.08}$ & $-3.9$ \\
    NGC 5765B & 126.3 & 95 & (c) & 7.67 & 0.33 & 1.20 & 7.65 & $-1.97^{+0.11}_{-0.11}$ & $44.34^{+0.16}_{-0.26}$ & $-1.4$\\
    NGC 6264 & 144 & 90 & (d) & 7.46 & 0.27 & 0.48 & 4.94 & $-1.74^{+0.11}_{-0.14}$ & $44.75^{+1.27}_{-0.60}$ & $-0.8$\\
    NGC 6323 & 107 & 89 & (e) & 6.99 & 0.15 & 0.31 & 1.67 & $-2.32^{+0.22}_{-0.36}$ & $45.20^{+0.14}_{-0.31}$ & $+0.1$\\
    UGC 3789 & 49.6 & 91 & (f) & 7.03 & 0.08 & 0.20 & 4.01 & $-3.15^{+0.20}_{-0.24}$ & $43.66^{+0.49}_{-0.25}$ & $-1.5$ \\
\hline    
\end{tabular}
\tablefoot{General properties of the sources considered. (1) -- Name of the megamaser galaxy. (2) -- Geometric distance in Mpc. (3) -- Inclination angle of the maser disk at its innermost radius, rounded to the nearest integer. (4) -- Reference for the geometric distance and for the inclination angle: (a) -- \citet{impellizzeri12}; (b) -- \citet{humphreys13}; (c) -- \citet{gao16}; (d) -- \citet{kuo13}; (e) -- \citet{kuo15}; (f) -- \citet{reid13}. (5) -- Logarithm of the black hole mass, in solar masses. (6-7) -- Inner and outer masing radii, respectively, in pc. (8) -- Radius of the sphere of influence of the BH, computed as $R_{\rm infl} = GM/\sigma^2$. The velocity dispersions $\sigma$ are taken from \citet{greene16}. All maser disks considered here are well within the sphere of influence of the BH, which dominates the gravitational potential. (9) -- Logarithm of the BH accretion rate in solar masses per year. The reference for columns 5, 6, 7 and 9 is \citet{kuo18}. (10) -- Logarithm of the bolometric luminosity in \lumcgs, computed as $L_{\rm bol} = 20 L^{\rm int}_{2-10}$
(see text). (11) -- Logarithm of the Eddington ratio, computed as $L_{\rm bol}/L_{\rm Edd}$, where $L_{\rm Edd} = 1.26\times10^{38}M_\bullet/M_\odot$.}
\end{table*}

\section{Basic Assumptions and Methodology}\label{sec:methodology}
When gas is accreted onto a BH, it is believed to form a disk whose angular momentum vector, at first order, aligns with the spin axis of the BH. In the inner disk regions the dynamics is regulated by the gravity of the massive, rotating BH, and the spacetime metric is described by the Kerr metric \citep{kerr63}. The method devised by \citet{campitiello18} relies on the numerical model \kerrbb \citep{li05}, which is implemented in XSPEC \citep{arnaud96}, though originally built for stellar mass BHs. This  describes the emission from a thin, steady state, general relativistic accretion disk around a rotating Kerr BH. It takes into account all the relativistic effects (i.e. Doppler beaming, gravitational redshift, light bending, self-irradiation, limb darkening), as well as the effects related to the black hole spin $a$ that determines the innermost stable circular orbit (ISCO), which in turn regulates the radiative efficiency of the system. \citet{campitiello18} derived analytical formulae linking observables to the physical parameters. In particular, given a spin value $a$, mass \mbh and accretion rate \mdot, for a system observed at an angle $\theta$, the frequency at which its accretion disk SED peaks $\nu_p$, and its intensity $(\nu L_\nu)_p$ can be expressed as
\begin{subequations}
\begin{eqnarray}
    \nu_p = \mathcal{A}\dot{M}^{1/4}M_{\bullet,9}^{-1/2} \cdot g_1(a,\theta) \label{eqn:2a} \\
    (\nu L_\nu)_p = \mathcal{B}\dot{M}\cos{\theta} \cdot g_2(a,\theta) \label{eqn:2b},
\end{eqnarray}
\end{subequations}
where $\log{\mathcal{A}} = 15.25$, $\log{\mathcal{B}} = 45.66$ \citep{calderone13}, \mdot is measured in \msun yr$^{-1}$, $M_{\bullet,9}$ in units of $10^9$ \msun, and the functions $g_1(a,\theta)$ and $g_2(a,\theta)$ encode the spin dependency of the observables, and can be numerically computed with \kerrbb, as explained below. 

If, for any reason, the optical/UV SED of a given source is not observable or available, however, we can focus on Equation \eqref{eqn:2b} alone, and indirectly link the left-hand side of the equation to the bolometric luminosity of the system. First, we consider that $L_{\rm bol} \approx 2L^{\rm obs}_{\rm d}$ \citep{calderone13}, where $L^{\rm obs}_{\rm d}$ is the observed accretion disk luminosity. Due to the angle-dependent emitted radiation pattern in the Kerr metric, the \textit{total} and \textit{observed} accretion disk luminosities differ: $L^{\rm obs}_{\rm d} = f(a,\theta)L_{\rm d}$, where $f(a,\theta)$ encodes the spin and angular dependence of the emission. Throughout this work, its functional form and coefficients are adopted as reported in Table B.1 of \citet{campitiello18}. Second, we take advantage of the self-similarity of accretion disk spectra, which implies the existence of relations among quantities at the peak frequencies, and the disk luminosity $L_{\rm d}$. In particular, $(\nu L_\nu)_p \approx L_{\rm d}/2$ \citep[see Eq. A10 of][]{calderone13}. Thus, 
\begin{equation} \label{eqn:6}
    (\nu L_\nu)_p \approx L_{\rm d}/2 = \frac{L^{\rm obs}_{\rm d}}{2f(a,\theta)} \approx \frac{L_{\rm bol}}{4f(a,\theta)}.
\end{equation} 

Hence, we can combine Equations \eqref{eqn:2b} and \eqref{eqn:6}, and express $L_{\rm bol}$ as a function of the other parameters:
\begin{equation} \label{eqn:7}
\begin{split}
    L_{\rm bol} & = 4\mathcal{B}\dot{M}\cos{\theta}\, g_2(a,\theta) f(a,\theta).
\end{split}
\end{equation}
This last equation shows that, under the assumption that the bolometric luminosity is a proxy for the accretion disk peak luminosity, if one can independently estimate the bolometric luminosity, the accretion rate, and the inclination angle, it is possible to infer the spin of the BH, even if the SED emission of the BBB is not directly observed, since the functions $f$ and $g_2$ are functions of the spin alone once $\theta$ is known. More specifically, this can be done by comparing the predicted $L_{\rm bol}(a)$ given by Equation \ref{eqn:7} with the actual estimated value.

\section{Sample selection}\label{sec:sample}

To successfully test Equation \ref{eqn:7} in constraining the spin without any optical/UV SED fitting, three observables are needed: the inclination angle of the system, its accretion rate and its bolometric luminosity. As already mentioned, disk megamasers are an ideal test case, given their precise almost edge-on geometry which allows us to get a precise handle on their inclination angle.
Moreover, the very fact that masers are detected implies very large obscuration in both the optical and X-ray band, often above the Compton-thick threshold \citep[$N_{\rm H} \gtrsim 10^{24}$ cm$^{-2}$,][]{masini16,masini19} -- hence the optical SED is completely dominated by the host stellar light, preventing a measurement of both the peak frequency $\nu_{\rm p}$ and intensity $(\nu L_\nu)_p$ of their accretion disks. 

Estimating the accretion rate \mdot of obscured AGN is very difficult and model dependent, and not all disk megamasers have an estimate available. \citet{kuo18} presented a sub-sample of six disk megamasers (NGC 2960, NGC 4258, NGC 5765B, NGC 6264, NGC 6323, and UGC 3789) for which an indirect estimate of the black hole accretion rate \mdot has also been derived. Specifically, the accretion rate has been measured with the goal of assessing the relative importance of the maser disk mass in the BH mass determination \citep{kuo18}, adopting the model from \citet{herrnstein05}. This model assumes a steady state, \citet{shakurasunyaev73} accretion disk extending out to the radii where the masers are produced. At each maser spot location, the Keplerian velocity is dictated by the total (BH + disk) enclosed mass at that radius:
\begin{equation} \label{eqn:8}
    M_{\rm tot}(r) = M_\bullet + 8.3\times10^4 \left(\frac{\dot{M}}{\alpha} \frac{M^{1/2}_\bullet(r^{1/2}-R^{1/2}_{\rm in})}{c^2_s}\right) M_\odot,
\end{equation}
where $\alpha$ is the \citet{shakurasunyaev73} viscosity parameter, $c^2_s$ is the sound speed, assumed to be $c^2_s = 2.15 \pm 0.15$ km s$^{-1}$, corresponding to a temperature suitable for maser emission ($T = 700 \pm 100$ K), and $R_{\rm in}$ is the innermost maser radius. By fitting the Keplerian rotation curve $v(r)$ of the maser spots, \citet{kuo18} obtain $M_{\rm tot}(r)$, from which the accretion rate is derived. Since no assumption is made over the viscosity parameter $\alpha$, the inferred accretion rates are actually $\dot{M}/\alpha$: we shall discuss this caveat later.
Hence, the only missing ingredient to derive the spin is the bolometric luminosity of the AGN in the \citet{kuo18} sample, $L_{\rm bol}$.

Because of the paucity of information on the above quantities (whose uncertainties will be discussed later), the selected sample is ideal to assess the validity of the proposed approach.

One implicit assumption in selecting this particular sample is that the maser disk is well within the sphere of influence of the SMBH. We check this assumption by investigating how each maser disk size compares with respect to the gravitational sphere of influence of its central SMBH. The radius of the sphere of influence is defined as $r_{\rm infl} = GM_\bullet/\sigma^2$, where $\sigma$ is the stellar velocity dispersion in the nuclear region of each galaxy, as reported by \citet{greene16}. Table \ref{tab:info} shows that the outer maser radii of the sources in our sample are always well within the radii of influence of their respective SMBHs, which span roughly an order of magnitude ($r_{\rm infl} \sim 1.9-13$ pc). Hence, the dynamics is dominated by the gravity of the SMBH.

In the following, we briefly present the properties and the (indirect) estimates of the luminosity of the sources. The bolometric luminosity is derived from the X-ray coronal luminosity by applying a single bolometric correction $k_{\rm bol} = 20$, appropriate for the range of X-ray luminosity of our sources \citep{lusso12, duras20}. We will compare our X-ray-based bolometric luminosities with those derived through the [OIII]$\lambda5007$ forbidden optical line ([OIII] hereafter), as reported by \citet{kuo20}. Details on our own spectral analysis are given in Appendix \ref{sec:appendix}, while useful information about the sample is summarized in Table \ref{tab:info}.

\subsection{NGC 2960}
NGC 2960 is a nearby spiral galaxy \citep[$z=0.01645$,][]{3rc91} whose nucleus hosts a SMBH of \mbh $= 1.16 \pm 0.05 \times 10^7$ \msun. The maser disk, discovered by \citet{henkel02}, is highly inclined \citep{kuo11, impellizzeri12} and likely responsible for the large obscuration affecting the nuclear emission. In the X-ray band, a short \nustar snapshot (20 ks) is presented by \citet{masini16}, pointing to high obscuration but inconclusive about the Compton-thick nature of the AGN. A re-analysis of the broadband ($0.2-70$ keV) \xmm + \nustar spectrum with the most up-to-date toroidal modeling gives a bolometric luminosity $L_{\rm bol} \sim 1.3^{+1.0}_{-0.5} \times 10^{43}$ \lumcgs.

\subsection{NGC 4258}
NGC 4258, at a distance of $D \sim 7.6$ Mpc \citep{humphreys13} is considered the archetypal disk megamaser, discovered almost three decades ago and extremely well studied ever since \citep{miyoshi95,gammie99,herrnstein05,humphreys13}. Being among the closest disk megamasers, it has been observed at almost all wavelengths. Despite its uniqueness, it is not the most representative example of the disk megamaser class, being only moderately obscured in the X-ray band \citep[$\lesssim 10^{23}$ cm$^{-2}$,][]{fruscione05}, and possibly powered by a radiatively inefficient accretion flow \citep{lasota96,herrnstein98,gammie99,yuan02}. Different papers point toward a bolometric luminosity $L_{\rm bol} \sim 10^{42}$ \lumcgs; here, we adopt an X-ray derived $L_{\rm bol} \sim 6 \pm 1 \times 10^{41}$ \lumcgs \citep[Masini et al. in prep.]{fruscione05}.

\subsection{NGC 5765B}
NGC 5765B is part of an merging pair of late type galaxies at $z=0.02754$ \citep{ahn12}. The megamaser disk is extensively studied in \citet{gao16}, and it is the largest of the sample considered here, extending out to $\sim 1.2$ pc from the nucleus. The broadband X-ray spectrum and a discussion on the AGN bolometric luminosity and Eddington ratio are presented by \citet{masini19}, and we adopt here an X-ray derived $L_{\rm bol} \sim 2.2^{+1.0}_{-1.0} \times 10^{44}$ \lumcgs.

\subsection{NGC 6264}
NGC 6264 is the farthest galaxy in the sample, at a distance of $D \sim 144$ Mpc \citep{kuo13}. Its central AGN is likely very obscured, with an obscuring column density over the Compton-thick threshold as measured with \xmm data alone \citep{castangia13}. Our own re-analysis of the \xmm spectrum gives $L_{\rm bol} \sim 5.6^{+98.0}_{-4.2} \times 10^{44}$ \lumcgs. 

\subsection{NGC 6323}
The maser spots in the nucleus of NGC 6323 \citep[$D \sim 107$ Mpc,][]{kuo15} trace an edge on Keplerian disk orbiting around a $M_\bullet = 9.4^{+3.7}_{-2.6} \times 10^6$ \msun SMBH \citep{kuo15}. To obtain an estimate of the bolometric luminosity  we analyzed an archival \xmm snapshot of $\sim 20$ ks. The tentative inferred value is $L_{\rm bol} \sim 1.6^{+0.6}_{-0.8} \times 10^{45}$ \lumcgs.

\subsection{UGC 3789}
Until the relatively recent discovery of water megamasers in its nucleus, UGC 3789 \citep[$D \sim 49.6$ Mpc,][]{reid13} was not known to host an AGN \citep{braatzgugliucci08}. Later on, UGC 3789 became a key galaxy to measure and refine the Hubble constant measurement using its masers \citep{reid09,braatz10,reid13}. Similarly to NGC 2960 and NGC 6323, there are no public optical spectra from which the O[III] line could be used to infer the bolometric luminosity. On the other hand, similarly to NGC 6264, \citet{castangia13} found evidence of heavy obscuration (possibly Compton-thick) using \xmm data. A re-analysis of the \xmm spectrum returns a bolometric luminosity $L_{\rm bol} = 4.6^{+9.6}_{-2.0} \times 10^{43}$ \lumcgs.

\section{Results}\label{sec:results}

\begin{figure*}
   \centering
   \includegraphics[width=\textwidth]{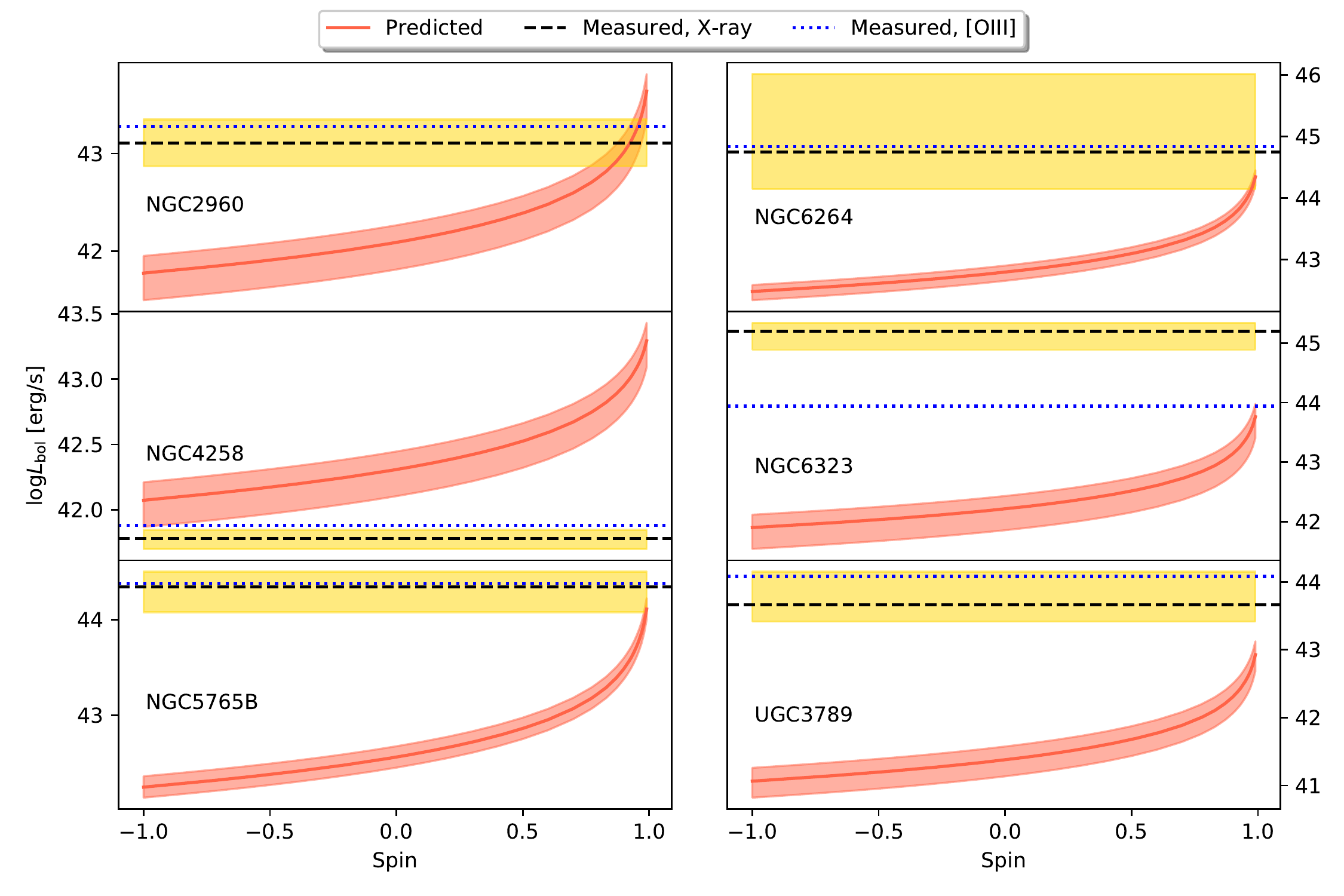}
   \caption{Bolometric luminosity -- spin plane for the six megamasers considered in this work. In each panel, the red line and area mark the expected $L_{\rm bol}$ based on Equation \eqref{eqn:7}, while the value as adopted or measured by us is marked by the dashed black line (and its $1\sigma$ uncertainty in yellow). We also plot the bolometric luminosity as measured from the [OIII] line \citep{kuo20} as dotted blue lines.}  In two cases (NGC 4258, UGC 3789) there is no formal solution between the expected and measured luminosities, while for NGC 6323 the two luminosities differ by more than an order of magnitude, suggesting that the X-ray intrinsic luminosity might be overestimated. Large spin values are generally preferred, although the results are tentative due to the uncertainties involved. Note the different scale on the $y$-axis in each panel.
   \label{fig:spins}
\end{figure*}

Once $\theta$ and \mdot are known, we need to compute numerically the function $g_2$ through \kerrbb to get the expected $L_{\rm bol}$ as a function of spin, to be compared with the values we have measured in the previous Section. In using \kerrbb, we assume a standard Keplerian disk with zero torque at the inner boundary (i.e., parameter \texttt{eta} of \kerrbb is fixed to zero), and we include the effects of self-irradiation and limb darkening (by setting the \kerrbb \texttt{rflag} and \texttt{lflag} parameters, respectively, to one). Then, \kerrbb is run over a grid of spin values; using Equation \eqref{eqn:2b} we derive the function $g_2$ for a given pair of (\mbh, \mdot) adopted in XSPEC. The inclination angle $\theta$ is always fixed to 85\degree, which is the model maximal allowed value \citep[except  for NGC 4258, for which $\theta = 72\degree$ is reported,][]{humphreys13}. We do not expect this to significantly impact on our conclusions for the range of observed inclination angles of our sources, but we discuss more this point in \S \ref{sec:discussion}.

The uncertainty on the \mdot largely affects the spread over the expected $L_{\rm bol}$, which translates in an uncertainty over the spin value when compared with the measured bolometric luminosity. As shown in Figure \ref{fig:spins}, the expected bolometric luminosity increases with increasing spin, as does the radiative efficiency. At face value, the results point toward generally large spin values, owing to the large uncertainty over the luminosity. A unique behaviour is observed for NGC 4258, which is arguably the most peculiar disk megamaser, presenting the lowest inclination angle, the lowest obscuration along the line of sight, and the lowest accretion rate in Eddington units (see Table \ref{tab:info}), likely in the RIAF regime (where we would not expect the assumptions underlying \kerrbb to be valid\footnote{We note that a larger bolometric correction ($k_{\rm bol} > 20$) would be needed to shift up the measured $L_{\rm bol}$, allowing a formal solution with the predicted curve. However, this possibility seems unlikely given the known scaling of $k_{\rm bol}$ with X-ray luminosity \citep{netzer19, duras20}}). For NGC 6323 and UGC 3789 instead there is no formal agreement between the expected luminosities and their X-ray measured ones. On the other hand, NGC 2960, NGC 5765B and NGC 6264 are formally consistent with a maximal spin value. 

\section{Discussion}\label{sec:discussion}
The results presented in the previous Section are inevitably tentative at this stage, given  the significant caveats and limitations, which we discuss in the following.

First of all, it is worth stressing that the uncertainty over the measured $L_{\rm bol}$ is inferred from the intrinsic X-ray luminosity alone; i.e., we have assumed a single bolometric correction $k_{\rm bol}=20$, suitable for the Seyfert-like luminosity of our sources \citep{lusso12,duras20}. The uncertainty over the intrinsic X-ray luminosity has been estimated from that of the normalization of the coronal power law of the X-ray spectroscopic analysis (Appendix \ref{sec:appendix}). Including the scatter on the correlation between the X-ray and the bolometric luminosity would broaden the yellow stripes in Figure \ref{fig:spins} by a factor of at least two.
However, we have also plotted in Figure \ref{fig:spins} all the [OIII]-derived bolometric luminosities as reported by \citet{kuo20}: they are fully consistent with the X-ray-derived bolometric luminosities in all cases, but for NGC 6323. For this last source, the [OIII]-derived bolometric luminosity suggests that the intrinsic X-ray luminosity of this AGN might be overestimated, and adopting the lower bolometric luminosity suggests a maximal spin value for NGC 6323 as well.

Furthermore, accurately measuring the inclination angle is very important for the successful application of this kind of analysis. In the previous Section we have implicitly assumed that the maser and inner accretion disks are co-planar. However, in principle their inclinations could significantly differ if some strong warping inside of the inner masing radius is present. This possibility can be tested by measuring the difference in position angle (PA) between the nuclear jets and the disk itself. Of the six sources in our sample, three have a detection of either pc or kpc-scale jets, and they are all consistent with being co-planar to the inner accretion disk. Indeed, \citet{herrnstein05} show that the core radio emission at 22 GHz is perpendicular to the maser disk in NGC 4258, and \citet{greene13} show that the PA difference between the maser disk and the kpc-scale jets in UGC 3789 and NGC 2960 are both $\sim80\degree-90\degree$. We note that \citet{kamali19} find that the jet in NGC 2960 is misaligned about 70\degree\xspace with respect to the masers: in this last case (i.e. assuming an inclination of 65\degree\xspace instead of 85\degree\xspace for NGC 2960), we find that its spin becomes consistent with lower values ($a<0.8$).
For the other three sources considered here, there is no jets detection nor radio continuum morphology reported in the literature, but \citet{greene13} demonstrated that megamaser disks are generally perpendicular to jets, thereby suggesting that the inclinations of the maser and the inner accretion disks are similar.

The accretion rate as reported by \citet{kuo18} and in the model of \citet{herrnstein05} is defined as $\dot{M}/\alpha$, where $\alpha$ is the viscosity parameter of the \citet{shakurasunyaev73} accretion disk model. A typical value for $\alpha$ found in numerical simulations of the inner regions of accretion disks and from theoretical considerations is in the range $\sim 0.01-0.4$ \citep{king07}, while here we implicitly assume that $\dot{M}/\alpha = \dot{M}$ at the inner maser radius (i.e., $\alpha=1$). It is unlikely that $\dot{M}/\alpha$ measured at the inner maser radius (i.e., at $\sim 10^5$ $r_{\rm g}$) is comparable with that close to the ISCO, unless the steady state condition of the disk is satisfied for a viscous timescale of $t_{\rm v} \sim 10^9$ yr \citep[e.g.,][]{gammie99}.
However, although the state of the art numerical simulations do not trace the evolution of such quantities at large distance from the SMBH, it is not unreasonable to think that both the accretion rate (e.g., due to variability and/or mass loss through winds) and the viscosity parameter should decrease getting closer to the BH horizon \citep{penna13}. If so, their ratio could in principle be consistent with that measured at the inner maser radius.

Very recently, \citet{ginerloeb21} proposed to estimate the spins of a sample of disk megamasers from the absence of Lense-Thirring precession \citep{lensethirring18} at the inner maser radius. Three of their sources are in common with ours. While we find consistent results for the spin of NGC 5765B, we are not able to set meaningful constraints for both NGC 4258 and UGC 3789. 
While the lack of constraints for NGC 4258 might be expected given the likely low radiatively efficient nature of its accretion flow \citep{yuan02}, it is not clear the reason for a lack of solution for UGC 3789 \citep[and similarly for NGC 6323, which is however not in the sample of ][]{ginerloeb21}. The bolometric luminosity might be overestimated (in particular for NGC 6323 as mentioned earlier, which results super-Eddington, as shown in Table \ref{tab:info}) if it were the column density, but we consider this possibility unlikely given the strong obscuration signatures in the X-ray spectra of the vast majority of disk megamasers. The observed discrepancy could lie in the mentioned unknown value of $\alpha$ at large scales. If the issue were related to the adopted bolometric correction, a simple estimate shows that the assumption of a bolometric correction of 10 instead of 20 would make UGC 3789 barely consistent with the most extreme prograde spin value, shifting up the measured $L_{\rm bol}$ of $\sim 0.25$ dex.

\section{Conclusions}\label{sec:conclusions}
Here we propose a method to constrain the spin of obscured AGN, specifically disk megamasers. This relies on the fact that in such sources the BH mass and inclination are well determined, and the accretion rate has been estimated through the dynamics of the maser spots. We explored both its theoretical and observational applicability.
We stress that the work presented intends to propose the new approach and procedure \citep[somehow alternative and complementary to the approach proposed by][]{ginerloeb21}, rather than effectively determining individual spins.
In particular, we first adapted the equations for sources where the accretion disk emission is not observed in the broadband SED. Then, we selected a sample of six well known local obscured AGN for which the accretion rate has been estimated through dynamical fitting of their water maser emission. Comparing the expected bolometric luminosities with the estimated values resulted in three (possibly four, considering the bolometric luminosities measured from the [OIII] line) sources being consistent with large spin values, while in other two no constraints can be set. The results, although tentative, suggest that the method could be successfully applied to obscured AGN, provided the underlying assumptions (most importantly the role and meaning of the accretion disk viscosity parameter) are explored and tested.

It is worth noticing that more stringent results could be obtained by directly detecting the BBB signature in the SED using polarized light \citep[e.g.,][]{antonuccimiller85}.

\begin{acknowledgements}
We thank the referee for an incredibly useful and kind report. Their suggestions made this paper stronger and clearer.
This research has made use of observations obtained with \xmm, an ESA science mission with instruments and contributions directly funded by ESA Member States and NASA. This work made also use of data from the \nustar mission, a project led by the California Institute of Technology, managed by the Jet Propulsion Laboratory, and funded by the National Aeronautics and Space Administration, as well as the \nustar Data Analysis Software (NuSTARDAS) jointly developed by the ASI Science Data Center (ASDC, Italy) and the California Institute of Technology (USA).
\end{acknowledgements}

\bibliographystyle{aa} 
\bibliography{aa2021-42451} 

\begin{thebibliography}{62}
\expandafter\ifx\csname natexlab\endcsname\relax\def\natexlab#1{#1}\fi

\bibitem[{{Ahn} {et~al.}(2012){Ahn}, {Alexandroff}, {Allende Prieto},
  {Anderson}, {Anderton}, {Andrews}, {Aubourg}, {Bailey}, {Balbinot}, {Barnes},
  \& et~al.}]{ahn12}
{Ahn}, C.~P., {Alexandroff}, R., {Allende Prieto}, C., {et~al.} 2012, \apjs,
  203, 21

\bibitem[{{Antonucci} \& {Miller}(1985)}]{antonuccimiller85}
{Antonucci}, R.~R.~J. \& {Miller}, J.~S. 1985, \apj, 297, 621

\bibitem[{{Arnaud}(1996)}]{arnaud96}
{Arnaud}, K.~A. 1996, in Astronomical Society of the Pacific Conference Series,
  Vol. 101, Astronomical Data Analysis Software and Systems V, ed. G.~H.
  {Jacoby} \& J.~{Barnes}, 17

\bibitem[{{Balokovi{\'c}} {et~al.}(2018){Balokovi{\'c}}, {Brightman},
  {Harrison}, {Comastri}, {Ricci}, {Buchner}, {Gandhi}, {Farrah}, \&
  {Stern}}]{balokovic18}
{Balokovi{\'c}}, M., {Brightman}, M., {Harrison}, F.~A., {et~al.} 2018, \apj,
  854, 42

\bibitem[{{Blandford} \& {Znajek}(1977)}]{blandfordznajek77}
{Blandford}, R.~D. \& {Znajek}, R.~L. 1977, \mnras, 179, 433

\bibitem[{{Braatz} \& {Gugliucci}(2008)}]{braatzgugliucci08}
{Braatz}, J.~A. \& {Gugliucci}, N.~E. 2008, \apj, 678, 96

\bibitem[{{Braatz} {et~al.}(2010){Braatz}, {Reid}, {Humphreys}, {Henkel},
  {Condon}, \& {Lo}}]{braatz10}
{Braatz}, J.~A., {Reid}, M.~J., {Humphreys}, E.~M.~L., {et~al.} 2010, \apj,
  718, 657

\bibitem[{{Buchner} {et~al.}(2015){Buchner}, {Georgakakis}, {Nandra},
  {Brightman}, {Menzel}, {Liu}, {Hsu}, {Salvato}, {Rangel}, {Aird}, {Merloni},
  \& {Ross}}]{buchner15}
{Buchner}, J., {Georgakakis}, A., {Nandra}, K., {et~al.} 2015, \apj, 802, 89

\bibitem[{{Calderone} {et~al.}(2013){Calderone}, {Ghisellini}, {Colpi}, \&
  {Dotti}}]{calderone13}
{Calderone}, G., {Ghisellini}, G., {Colpi}, M., \& {Dotti}, M. 2013, \mnras,
  431, 210

\bibitem[{{Campitiello} {et~al.}(2018){Campitiello}, {Ghisellini}, {Sbarrato},
  \& {Calderone}}]{campitiello18}
{Campitiello}, S., {Ghisellini}, G., {Sbarrato}, T., \& {Calderone}, G. 2018,
  \aap, 612, A59

\bibitem[{{Cash}(1979)}]{cash79}
{Cash}, W. 1979, \apj, 228, 939

\bibitem[{{Castangia} {et~al.}(2013){Castangia}, {Panessa}, {Henkel}, {Kadler},
  \& {Tarchi}}]{castangia13}
{Castangia}, P., {Panessa}, F., {Henkel}, C., {Kadler}, M., \& {Tarchi}, A.
  2013, \mnras, 436, 3388

\bibitem[{{de Vaucouleurs} {et~al.}(1991){de Vaucouleurs}, {de Vaucouleurs},
  {Corwin}, {Buta}, {Paturel}, \& {Fouque}}]{3rc91}
{de Vaucouleurs}, G., {de Vaucouleurs}, A., {Corwin}, Herold~G., J., {et~al.}
  1991, {Third Reference Catalogue of Bright Galaxies}

\bibitem[{{Duras} {et~al.}(2020){Duras}, {Bongiorno}, {Ricci}, {Piconcelli},
  {Shankar}, {Lusso}, {Bianchi}, {Fiore}, {Maiolino}, {Marconi}, {Onori},
  {Sani}, {Schneider}, {Vignali}, \& {La Franca}}]{duras20}
{Duras}, F., {Bongiorno}, A., {Ricci}, F., {et~al.} 2020, \aap, 636, A73

\bibitem[{{Ferrarese} \& {Merritt}(2000)}]{ferraresemerritt00}
{Ferrarese}, L. \& {Merritt}, D. 2000, \apjl, 539, L9

\bibitem[{{Fruscione} {et~al.}(2005){Fruscione}, {Greenhill}, {Filippenko},
  {Moran}, {Herrnstein}, \& {Galle}}]{fruscione05}
{Fruscione}, A., {Greenhill}, L.~J., {Filippenko}, A.~V., {et~al.} 2005, \apj,
  624, 103

\bibitem[{{Gammie} {et~al.}(1999){Gammie}, {Narayan}, \&
  {Blandford}}]{gammie99}
{Gammie}, C.~F., {Narayan}, R., \& {Blandford}, R. 1999, \apj, 516, 177

\bibitem[{{Gao} {et~al.}(2016){Gao}, {Braatz}, {Reid}, {Lo}, {Condon},
  {Henkel}, {Kuo}, {Impellizzeri}, {Pesce}, \& {Zhao}}]{gao16}
{Gao}, F., {Braatz}, J.~A., {Reid}, M.~J., {et~al.} 2016, \apj, 817, 128

\bibitem[{{Gebhardt} {et~al.}(2000){Gebhardt}, {Bender}, {Bower}, {Dressler},
  {Faber}, {Filippenko}, {Green}, {Grillmair}, {Ho}, {Kormendy}, {Lauer},
  {Magorrian}, {Pinkney}, {Richstone}, \& {Tremaine}}]{gebhardt00}
{Gebhardt}, K., {Bender}, R., {Bower}, G., {et~al.} 2000, \apjl, 539, L13

\bibitem[{{George} \& {Fabian}(1991)}]{georgefabian91}
{George}, I.~M. \& {Fabian}, A.~C. 1991, \mnras, 249, 352

\bibitem[{{Giner} \& {Loeb}(2021)}]{ginerloeb21}
{Giner}, S. \& {Loeb}, A. 2021, arXiv e-prints, arXiv:2104.05084

\bibitem[{{Gliozzi} {et~al.}(2021){Gliozzi}, {Williams}, \&
  {Michel}}]{gliozzi21}
{Gliozzi}, M., {Williams}, J.~K., \& {Michel}, D.~A. 2021, \mnras, 502, 3329

\bibitem[{{Greene} {et~al.}(2013){Greene}, {Seth}, {den Brok}, {Braatz},
  {Henkel}, {Sun}, {Peng}, {Kuo}, {Impellizzeri}, \& {Lo}}]{greene13}
{Greene}, J.~E., {Seth}, A., {den Brok}, M., {et~al.} 2013, \apj, 771, 121

\bibitem[{{Greene} {et~al.}(2016){Greene}, {Seth}, {Kim}, {L{\"a}sker},
  {Goulding}, {Gao}, {Braatz}, {Henkel}, {Condon}, {Lo}, \& {Zhao}}]{greene16}
{Greene}, J.~E., {Seth}, A., {Kim}, M., {et~al.} 2016, \apjl, 826, L32

\bibitem[{{Grier} {et~al.}(2017){Grier}, {Trump}, {Shen}, {Horne}, {Kinemuchi},
  {McGreer}, {Starkey}, {Brandt}, {Hall}, {Kochanek}, {Chen}, {Denney},
  {Greene}, {Ho}, {Homayouni}, {I-Hsiu Li}, {Pei}, {Peterson}, {Petitjean},
  {Schneider}, {Sun}, {AlSayyad}, {Bizyaev}, {Brinkmann}, {Brownstein},
  {Bundy}, {Dawson}, {Eftekharzadeh}, {Fernandez-Trincado}, {Gao},
  {Hutchinson}, {Jia}, {Jiang}, {Oravetz}, {Pan}, {Paris}, {Ponder}, {Peters},
  {Rogerson}, {Simmons}, {Smith}, \& {Wang}}]{grier17}
{Grier}, C.~J., {Trump}, J.~R., {Shen}, Y., {et~al.} 2017, \apj, 851, 21

\bibitem[{{Henkel} {et~al.}(2002){Henkel}, {Braatz}, {Greenhill}, \&
  {Wilson}}]{henkel02}
{Henkel}, C., {Braatz}, J.~A., {Greenhill}, L.~J., \& {Wilson}, A.~S. 2002,
  \aap, 394, L23

\bibitem[{{Herrnstein} {et~al.}(1998){Herrnstein}, {Greenhill}, {Moran},
  {Diamond}, {Inoue}, {Nakai}, \& {Miyoshi}}]{herrnstein98}
{Herrnstein}, J.~R., {Greenhill}, L.~J., {Moran}, J.~M., {et~al.} 1998, \apjl,
  497, L69

\bibitem[{{Herrnstein} {et~al.}(2005){Herrnstein}, {Moran}, {Greenhill}, \&
  {Trotter}}]{herrnstein05}
{Herrnstein}, J.~R., {Moran}, J.~M., {Greenhill}, L.~J., \& {Trotter}, A.~S.
  2005, \apj, 629, 719

\bibitem[{{Humphreys} {et~al.}(2013){Humphreys}, {Reid}, {Moran}, {Greenhill},
  \& {Argon}}]{humphreys13}
{Humphreys}, E.~M.~L., {Reid}, M.~J., {Moran}, J.~M., {Greenhill}, L.~J., \&
  {Argon}, A.~L. 2013, \apj, 775, 13

\bibitem[{{Impellizzeri} {et~al.}(2012){Impellizzeri}, {Braatz}, {Kuo}, {Reid},
  {Lo}, {Henkel}, \& {Condon}}]{impellizzeri12}
{Impellizzeri}, C.~M.~V., {Braatz}, J.~A., {Kuo}, C.-Y., {et~al.} 2012, in
  Cosmic Masers - from OH to H0, ed. R.~S. {Booth}, W.~H.~T. {Vlemmings}, \&
  E.~M.~L. {Humphreys}, Vol. 287, 311--315

\bibitem[{{Kalberla} {et~al.}(2005){Kalberla}, {Burton}, {Hartmann}, {Arnal},
  {Bajaja}, {Morras}, \& {P{\"o}ppel}}]{kalberla05}
{Kalberla}, P.~M.~W., {Burton}, W.~B., {Hartmann}, D., {et~al.} 2005, \aap,
  440, 775

\bibitem[{{Kamali} {et~al.}(2019){Kamali}, {Henkel}, {Koyama}, {Kuo}, {Condon},
  {Brunthaler}, {Reid}, {Greene}, {Menten}, {Impellizzeri}, {Braatz},
  {Litzinger}, \& {Kadler}}]{kamali19}
{Kamali}, F., {Henkel}, C., {Koyama}, S., {et~al.} 2019, \aap, 624, A42

\bibitem[{{Kerr}(1963)}]{kerr63}
{Kerr}, R.~P. 1963, \prl, 11, 237

\bibitem[{{King} {et~al.}(2007){King}, {Pringle}, \& {Livio}}]{king07}
{King}, A.~R., {Pringle}, J.~E., \& {Livio}, M. 2007, \mnras, 376, 1740

\bibitem[{{Kormendy} \& {Richstone}(1995)}]{kormendyrichstone95}
{Kormendy}, J. \& {Richstone}, D. 1995, \araa, 33, 581

\bibitem[{{Kuo} {et~al.}(2011){Kuo}, {Braatz}, {Condon}, {Impellizzeri}, {Lo},
  {Zaw}, {Schenker}, {Henkel}, {Reid}, \& {Greene}}]{kuo11}
{Kuo}, C.~Y., {Braatz}, J.~A., {Condon}, J.~J., {et~al.} 2011, \apj, 727, 20

\bibitem[{{Kuo} {et~al.}(2020){Kuo}, {Braatz}, {Impellizzeri}, {Gao}, {Pesce},
  {Reid}, {Condon}, {Kamali}, {Henkel}, \& {Greene}}]{kuo20}
{Kuo}, C.~Y., {Braatz}, J.~A., {Impellizzeri}, C.~M.~V., {et~al.} 2020, \mnras,
  498, 1609

\bibitem[{{Kuo} {et~al.}(2015){Kuo}, {Braatz}, {Lo}, {Reid}, {Suyu}, {Pesce},
  {Condon}, {Henkel}, \& {Impellizzeri}}]{kuo15}
{Kuo}, C.~Y., {Braatz}, J.~A., {Lo}, K.~Y., {et~al.} 2015, \apj, 800, 26

\bibitem[{{Kuo} {et~al.}(2013){Kuo}, {Braatz}, {Reid}, {Lo}, {Condon},
  {Impellizzeri}, \& {Henkel}}]{kuo13}
{Kuo}, C.~Y., {Braatz}, J.~A., {Reid}, M.~J., {et~al.} 2013, \apj, 767, 155

\bibitem[{{Kuo} {et~al.}(2018){Kuo}, {Constantin}, {Braatz}, {Chung},
  {Witherspoon}, {Pesce}, {Impellizzeri}, {Gao}, {Hao}, {Woo}, \&
  {Zaw}}]{kuo18}
{Kuo}, C.~Y., {Constantin}, A., {Braatz}, J.~A., {et~al.} 2018, \apj, 860, 169

\bibitem[{{Lasota} {et~al.}(1996){Lasota}, {Abramowicz}, {Chen}, {Krolik},
  {Narayan}, \& {Yi}}]{lasota96}
{Lasota}, J.~P., {Abramowicz}, M.~A., {Chen}, X., {et~al.} 1996, \apj, 462, 142

\bibitem[{{Lense} \& {Thirring}(1918)}]{lensethirring18}
{Lense}, J. \& {Thirring}, H. 1918, Physikalische Zeitschrift, 19, 156

\bibitem[{{Li} {et~al.}(2005){Li}, {Zimmerman}, {Narayan}, \&
  {McClintock}}]{li05}
{Li}, L.-X., {Zimmerman}, E.~R., {Narayan}, R., \& {McClintock}, J.~E. 2005,
  \apjs, 157, 335

\bibitem[{{Lusso} {et~al.}(2012){Lusso}, {Comastri}, {Simmons}, {Mignoli},
  {Zamorani}, {Vignali}, {Brusa}, {Shankar}, {Lutz}, {Trump}, {Maiolino},
  {Gilli}, {Bolzonella}, {Puccetti}, {Salvato}, {Impey}, {Civano}, {Elvis},
  {Mainieri}, {Silverman}, {Koekemoer}, {Bongiorno}, {Merloni}, {Berta}, {Le
  Floc'h}, {Magnelli}, {Pozzi}, \& {Riguccini}}]{lusso12}
{Lusso}, E., {Comastri}, A., {Simmons}, B.~D., {et~al.} 2012, \mnras, 425, 623

\bibitem[{{Magorrian} {et~al.}(1998){Magorrian}, {Tremaine}, {Richstone},
  {Bender}, {Bower}, {Dressler}, {Faber}, {Gebhardt}, {Green}, {Grillmair},
  {Kormendy}, \& {Lauer}}]{magorrian98}
{Magorrian}, J., {Tremaine}, S., {Richstone}, D., {et~al.} 1998, \aj, 115, 2285

\bibitem[{{Masini} {et~al.}(2016){Masini}, {Comastri}, {Balokovi{\'c}}, {Zaw},
  {Puccetti}, {Ballantyne}, {Bauer}, {Boggs}, {Brandt}, {Brightman},
  {Christensen}, {Craig}, {Gandhi}, {Hailey}, {Harrison}, {Koss}, {Madejski},
  {Ricci}, {Rivers}, {Stern}, \& {Zhang}}]{masini16}
{Masini}, A., {Comastri}, A., {Balokovi{\'c}}, M., {et~al.} 2016, \aap, 589,
  A59

\bibitem[{{Masini} {et~al.}(2019){Masini}, {Comastri}, {Hickox}, {Koss},
  {Civano}, {Brigthman}, {Brusa}, \& {Lanzuisi}}]{masini19}
{Masini}, A., {Comastri}, A., {Hickox}, R.~C., {et~al.} 2019, \apj, 882, 83

\bibitem[{{Merloni} {et~al.}(2014){Merloni}, {Bongiorno}, {Brusa}, {Iwasawa},
  {Mainieri}, {Magnelli}, {Salvato}, {Berta}, {Cappelluti}, {Comastri},
  {Fiore}, {Gilli}, {Koekemoer}, {Le Floc'h}, {Lusso}, {Lutz}, {Miyaji},
  {Pozzi}, {Riguccini}, {Rosario}, {Silverman}, {Symeonidis}, {Treister},
  {Vignali}, \& {Zamorani}}]{merloni14}
{Merloni}, A., {Bongiorno}, A., {Brusa}, M., {et~al.} 2014, \mnras, 437, 3550

\bibitem[{{Miyoshi} {et~al.}(1995){Miyoshi}, {Moran}, {Herrnstein},
  {Greenhill}, {Nakai}, {Diamond}, \& {Inoue}}]{miyoshi95}
{Miyoshi}, M., {Moran}, J., {Herrnstein}, J., {et~al.} 1995, \nat, 373, 127

\bibitem[{{Netzer}(2019)}]{netzer19}
{Netzer}, H. 2019, \mnras, 488, 5185

\bibitem[{{Penna} {et~al.}(2013){Penna}, {S{\k{a}}dowski}, {Kulkarni}, \&
  {Narayan}}]{penna13}
{Penna}, R.~F., {S{\k{a}}dowski}, A., {Kulkarni}, A.~K., \& {Narayan}, R. 2013,
  \mnras, 428, 2255

\bibitem[{{Reid} {et~al.}(2009){Reid}, {Braatz}, {Condon}, {Greenhill},
  {Henkel}, \& {Lo}}]{reid09}
{Reid}, M.~J., {Braatz}, J.~A., {Condon}, J.~J., {et~al.} 2009, \apj, 695, 287

\bibitem[{{Reid} {et~al.}(2013){Reid}, {Braatz}, {Condon}, {Lo}, {Kuo},
  {Impellizzeri}, \& {Henkel}}]{reid13}
{Reid}, M.~J., {Braatz}, J.~A., {Condon}, J.~J., {et~al.} 2013, \apj, 767, 154

\bibitem[{{Reines} \& {Volonteri}(2015)}]{reinesvolonteri15}
{Reines}, A.~E. \& {Volonteri}, M. 2015, \apj, 813, 82

\bibitem[{{Reynolds}(2019)}]{reynolds19}
{Reynolds}, C.~S. 2019, Nature Astronomy, 3, 41

\bibitem[{{Sesana} {et~al.}(2014){Sesana}, {Barausse}, {Dotti}, \&
  {Rossi}}]{sesana14}
{Sesana}, A., {Barausse}, E., {Dotti}, M., \& {Rossi}, E.~M. 2014, \apj, 794,
  104

\bibitem[{{Shakura} \& {Sunyaev}(1973)}]{shakurasunyaev73}
{Shakura}, N.~I. \& {Sunyaev}, R.~A. 1973, \aap, 24, 337

\bibitem[{{Shankar} {et~al.}(2009){Shankar}, {Weinberg}, \&
  {Miralda-Escud{\'e}}}]{shankar09}
{Shankar}, F., {Weinberg}, D.~H., \& {Miralda-Escud{\'e}}, J. 2009, \apj, 690,
  20

\bibitem[{{Soltan}(1982)}]{soltan82}
{Soltan}, A. 1982, \mnras, 200, 115

\bibitem[{{Tarchi}(2012)}]{tarchi12}
{Tarchi}, A. 2012, in IAU Symposium, Vol. 287, Cosmic Masers - from OH to H0,
  ed. R.~S. {Booth}, W.~H.~T. {Vlemmings}, \& E.~M.~L. {Humphreys}, 323--332

\bibitem[{{Volonteri} {et~al.}(2013){Volonteri}, {Sikora}, {Lasota}, \&
  {Merloni}}]{volonteri13}
{Volonteri}, M., {Sikora}, M., {Lasota}, J.~P., \& {Merloni}, A. 2013, \apj,
  775, 94

\bibitem[{{Yuan} {et~al.}(2002){Yuan}, {Markoff}, {Falcke}, \&
  {Biermann}}]{yuan02}
{Yuan}, F., {Markoff}, S., {Falcke}, H., \& {Biermann}, P.~L. 2002, \aap, 391,
  139

\end{thebibliography}

\appendix
\section{X-ray Spectroscopy}\label{sec:appendix}
In this Section, we briefly describe our own X-ray spectral analysis of both previously published and unpublished spectra. We used the same basic model components to be as homogeneous as possible; this baseline model is composed by an absorbed power law, reprocessing from the torus \citep[i.e., the \texttt{Borus} model,][]{balokovic18} -- which includes both cold reflection and fluorescence, and soft X-ray emission from a two-temperature hot plasma with possibly a scattered power law mirroring the coronal one. All these components are then absorbed by the Galactic column density, the amount of which has been estimated for each source through the \texttt{nh} command in XSPEC \citep{kalberla05}. In the XSPEC notation, the baseline model is implemented as follows:
\begin{multline}
    \overbrace{\texttt{tbabs}}^{\text{Galactic $N_{\rm H}$}}\times \{ \overbrace{\texttt{zphabs}\times \texttt{cabs}\times \texttt{cutoffpl}}^{\text{Intrinsic absorbed emission}} + \overbrace{\texttt{Borus02}}^{\text{Torus reprocessing}} + \\ + \underbrace{\texttt{mekal} + \texttt{mekal} + \texttt{const}\times \texttt{zpowerlw}}_{\text{Soft emission}}\}.
\end{multline}
When needed, a Gaussian line was added to the baseline model to fit the residuals at $\sim 1.7-1.8$ keV, which we interpreted as either the Si K$\alpha$ or K$\beta$ emission line.

The log of the observations used in this analysis is presented in Table \ref{tab:log}. The \xmm observations were reduced and products extracted with the \texttt{SAS} v1.3 software and relative standard tasks. The \nustar observations were reduced with the \texttt{nupipeline} v0.4.6 and \texttt{nuproducts} v0.3.0 tasks as part of the \texttt{NuSTARDAS} package. Unless otherwise specified, all the spectra have been rebinned to have at least 20 counts per bin. The best-fit values of the parameters are shown in Table \ref{tab:xraysp}. We notice that we could significantly constrain the photon index only for NGC 2960, thanks to the broadband \xmm + \nustar coverage. In the other cases, where only \xmm data are available, the photon index was unconstrained and was fixed to a common value of $\Gamma= 1.9$.

\subsection{NGC 2960}
NGC 2960 is firmly detected by \nustar up to $\sim 30$ keV. Its broadband ($0.3-78$ keV) spectrum was fit with the baseline model, with the addition of a cross-calibration \texttt{constant}, which takes into account both different calibration between the instruments, as as well as possible flux variations due to the non-simultaneity of the \xmm and \nustar observations. The spectrum is well fit ($\chi^2$/dof = 54/35) with the model, although few residuals can be seen at the Fe K$\alpha$ line energy at $6.4$ keV, which may indicate that the obscuration and reflection derived from the fit are too low to explain the line prominence.

\subsection{NGC 6264}
The \xmm spectrum is well fit by the baseline model. A residual at $\sim 1.8$ keV is consistent with the emission of the Si K$\beta$ line. The source is well within the Compton-thick regime of obscuration. In general, our parameters are consistent with those reported by \citet{castangia13}.

\subsection{NGC 6323}
NGC 6323 is the faintest source in our sample, and due to the limited spectral quality we employed the Cash statistic \citep{cash79} during the fit. Its \xmm spectrum is extremely flat at hard energies, showing a prominent Fe K$\alpha$ line. Both these features are typical of a strong reflection component, which is likely reflection-dominated. The best-fit baseline model requires indeed a low covering factor (CF $< 0.2$) to better fit the flat spectral emission at $2-5$ keV, and a very large column density as well. As a consequence, the intrinsic luminosity has to be very bright to produce the observed reprocessing features.

\subsection{UGC 3789}
Similarly to what found for NGC 6264, the \xmm spectrum is well fit by the baseline model. Residuals at $\sim 1.74$ keV  are consistent with the emission of the Si K$\alpha$ line. Further residuals at soft energies may suggest that a simple two temperature hot plasma is not suited to explain the emission. In general, the derived parameters are consistent with the ones reported by \citet{castangia13}.

\begin{table}
\caption{ObsIDs considered for the X-ray spectral analysis.}
\label{tab:log}
\centering                                    
\begin{tabular}{l l}          
\hline\hline                  
Source & ObsIDs \\
\hline
\multirow{2}{5em}{NGC 2960} & 0306050201 (\xmm) \\ 
& 60001069002 (\nustar) \\ 
NGC 6264 & 0654800201 (\xmm) \\ 
NGC 6323 & 0824970301 (\xmm) \\ 
UGC 3789 & 0654800101 (\xmm) \\
\hline
\end{tabular}
\end{table}

\begin{table*}
\caption{Results of the X-ray spectral analysis with the baseline model.}
\label{tab:xraysp}
\centering                                    
\begin{tabular}{l c c c c}          
\hline\hline                  
Parameters & NGC 2960 & NGC 6264 & NGC 6323 & UGC 3789 \\
\hline

    $\chi^2$/dof & 54/35 & 37/47 & 49/41 & 113/83 \\
    $\Gamma$ & $2.1^{+0.2}_{-0.3}$ & (1.9) & (1.9) & (1.9) \\
    $N_{\rm PL}$ [photons keV$^{-1}$ cm$^{-2}$] & $2.3^{+1.5}_{-0.9} \times 10^{-4}$ & $3.9^{+146.1}_{-2.9} \times 10^{-3}$ & $1.8^{+0.8}_{-0.4} \times 10^{-2}$ & $2.8^{+5.6}_{-1.3} \times 10^{-3}$ \\
    $\log{(N_{\rm H}/\text{cm$^{-2}$)}}$ & $23.84^{+0.11}_{-0.13}$ & $24.47^{+0.78}_{-0.18}$ & $25.38^{+u}_{-0.93}$ & $24.20^{+0.18}_{-0.09}$\\
    Covering Factor & (0.5) & (0.5) & $0.1^{+0.1}_{-l}$ & (0.5) \\
    $f_{\rm s}$ [\%] & $2.8^{+3.2}_{-1.4}$ & $0.1^{+0.3}_{-l}$ & $0.010^{+0.042}_{-0.006}$ & $0.5^{+0.4}_{-0.3}$ \\
    $E_{\text{line}}$ [keV] & $-$ & $1.80^{+0.03}_{-0.02}$ & $-$ & $1.74 \pm 0.03$ \\
    EW$_{\text{line}}$ [eV] & $-$ & $183^{+94}_{-93}$ & $-$ & $132^{+59}_{-61}$ \\
    $kT_1$ [keV] & $0.09^{+0.03}_{-l}$  & $0.12 \pm 0.01$ & $0.14^{+0.06}_{-0.04}$ & $0.115^{+0.005}_{-0.004}$ \\
    $N_{\rm mekal1}$ [photons keV$^{-1}$ cm$^{-2}$] & $2.4^{+2.5}_{-1.8} \times 10^{-5}$ & $4.4 \pm 1.1 \times 10^{-5}$ & $3.3^{+0.4}_{-1.9} \times 10^{-6}$ & $1.6 \pm 0.2 \times 10^{-4}$ \\
    $kT_2$ [keV] & $0.68^{+0.08}_{-0.07}$ & $0.63 \pm 0.03$ & $0.62^{+0.08}_{-0.12}$ & $0.64 \pm 0.02$ \\
    $N_{\rm mekal2}$ [photons keV$^{-1}$ cm$^{-2}$] & $3.4 \pm 0.9 \times 10^{-6}$ & $7.0 \pm 0.6 \times 10^{-6}$ & $2.4^{+0.4}_{-0.6} \times 10^{-6}$ & $2.20 \pm 0.14 \times 10^{-5}$\\
    $F^{\rm obs}_{2-10}$ [\fluxcgs] & $6.1 \times 10^{-14}$ & $3.1 \times 10^{-14}$ & $2.1 \times 10^{-14}$ & $1.6 \times 10^{-13}$ \\
    $F^{\rm int}_{2-10}$ [\fluxcgs] & $5.2 \times10^{-13}$ & $1.1 \times 10^{-11}$ & $4.7 \times 10^{-11}$ & $8.2 \times 10^{-12}$ \\
\hline
\end{tabular}
\tablefoot{To fit the \xmm spectrum of NGC 6323, we used the Cash statistic \citep{cash79} due to the low number of photons which were rebinned to be at least 3 counts per bin. All the other spectra were fitted with the $\chi^2$ statistic.}
\end{table*}

\begin{figure*}
   \centering
   \includegraphics[width=0.45\textwidth]{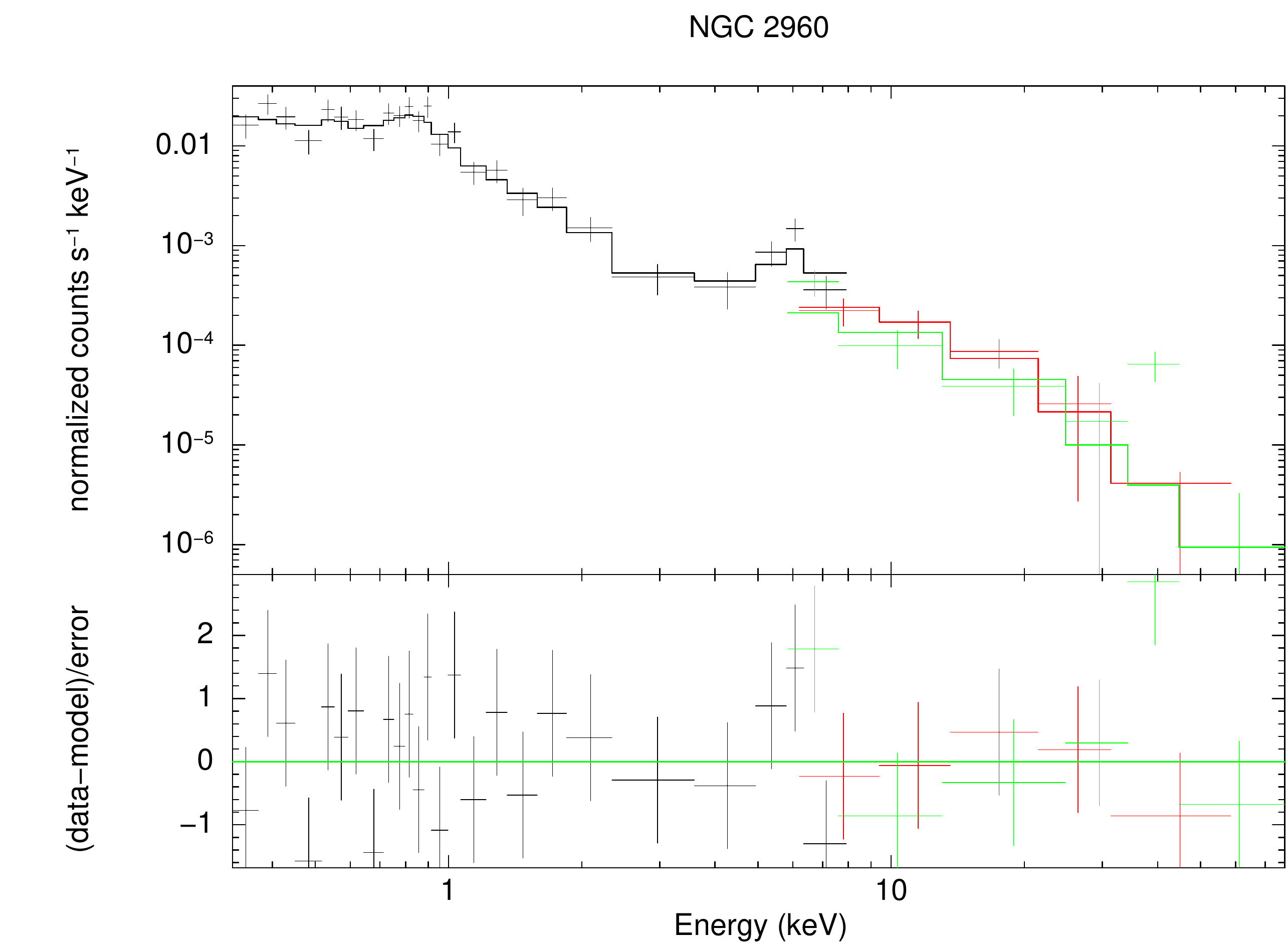}
   \includegraphics[width=0.45\textwidth]{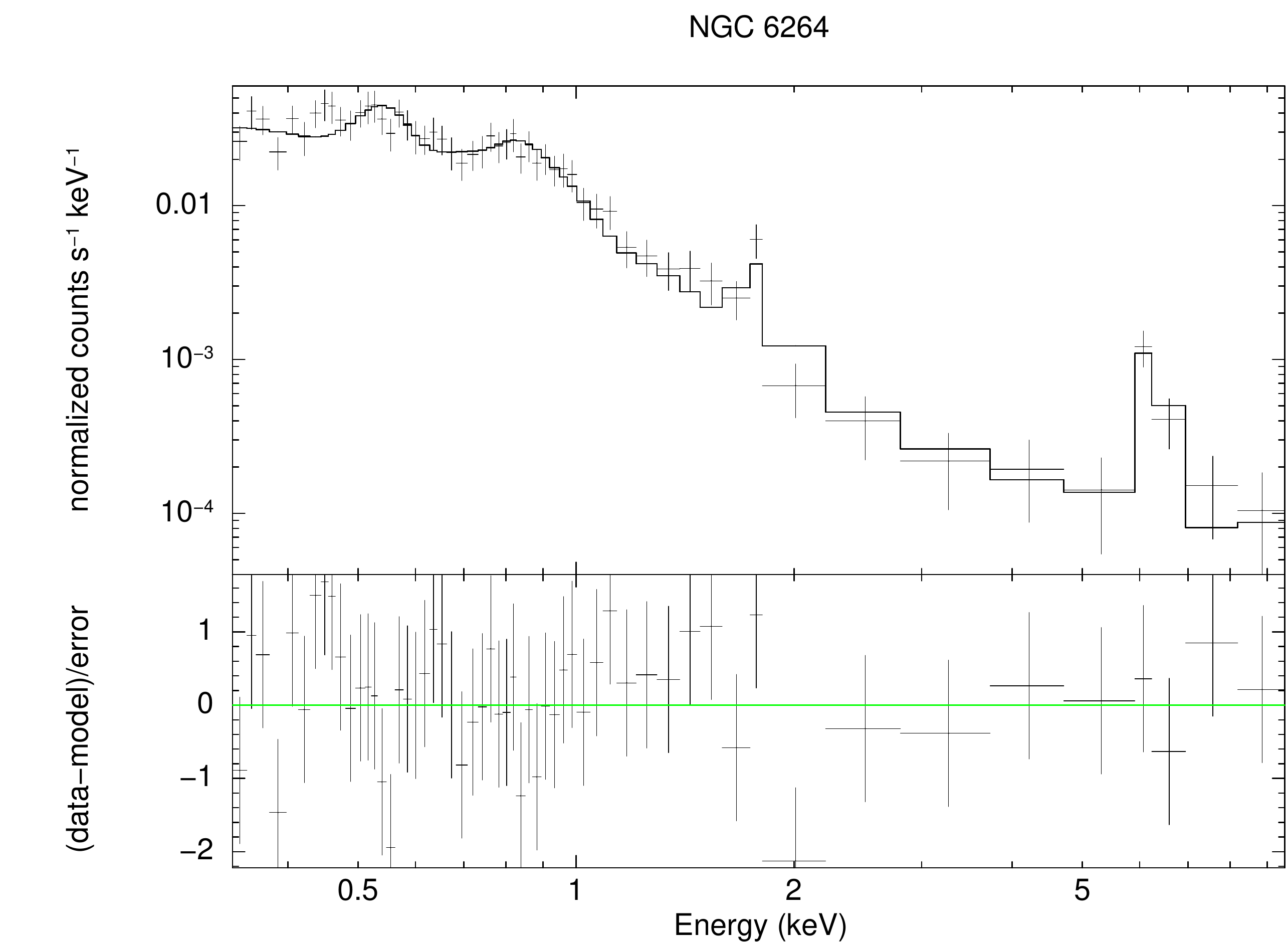}
   \includegraphics[width=0.45\textwidth]{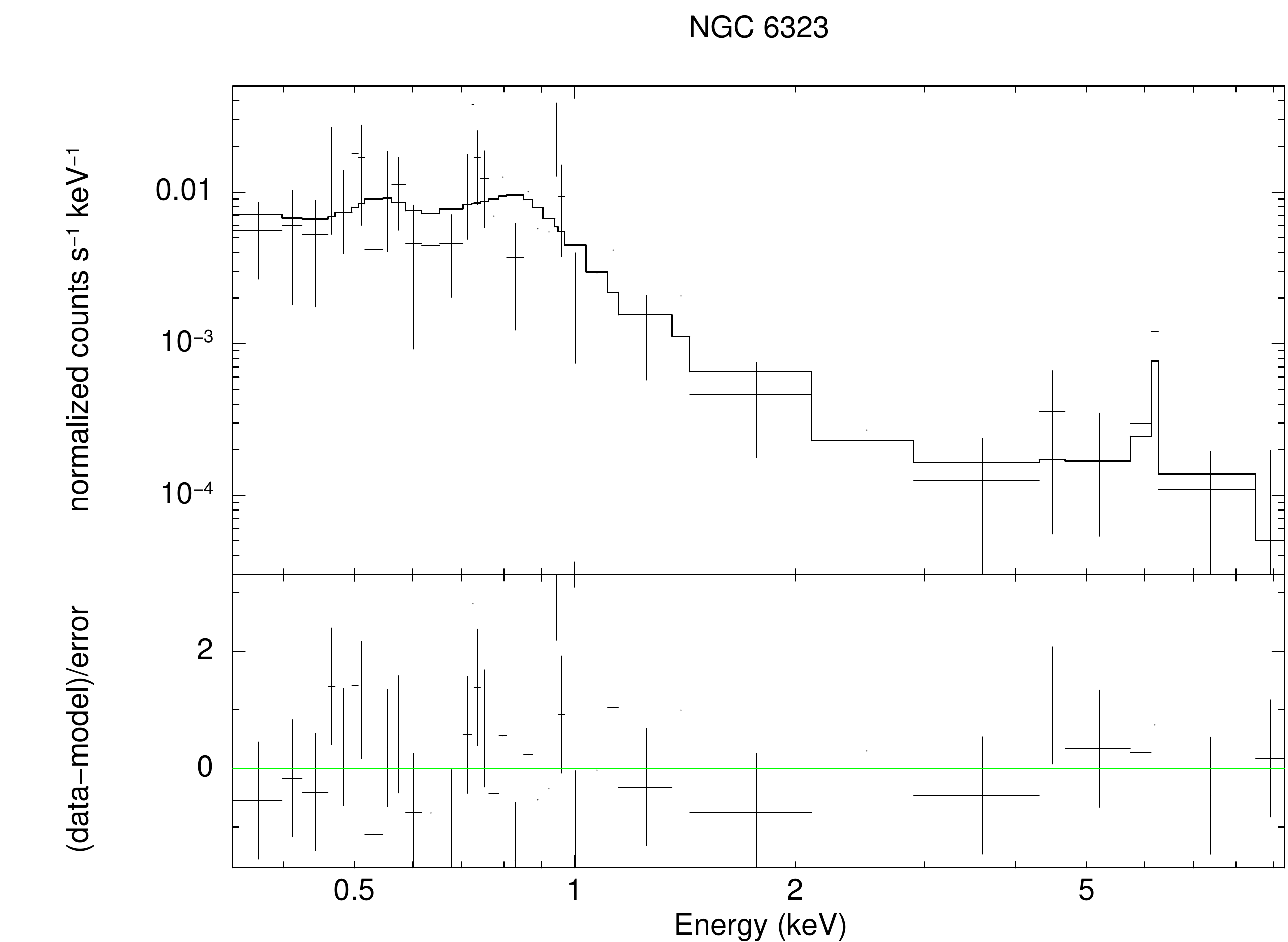}
   \includegraphics[width=0.45\textwidth]{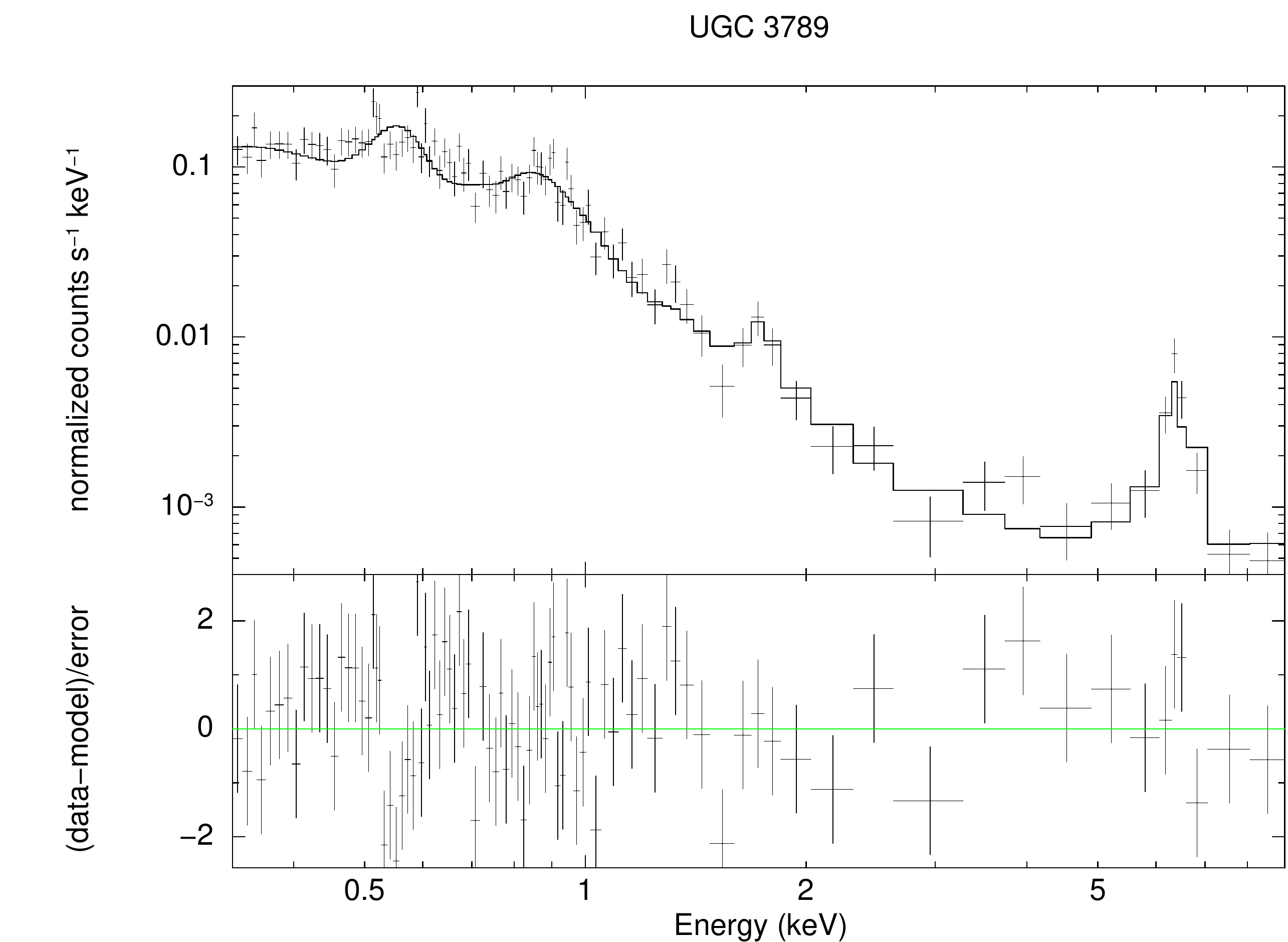}
   \caption{X-ray spectra fitted with the best-fit baseline model described in the text. In all the panels, the black line refers to \xmm PN data, while the red and green ones label \nustar FPMA and FPMB, respectively. All the spectra are heavily obscured, with prominent Fe K$\alpha$ emission lines.}
   \label{fig:xraysp}
\end{figure*}

\end{document}